\documentclass[10pt]{article}

\usepackage[margin=.75in]{geometry}

\usepackage{amsmath,amssymb}
\usepackage{graphicx}
\usepackage{subcaption}

\usepackage{color}
\definecolor{indigo}{RGB}{0,0,120}
\usepackage[colorlinks=true, linkcolor=indigo, citecolor=blue, urlcolor=indigo]{hyperref}

\newcommand{\Red}{\color{red}}

\newcommand{\beq}{\begin{equation}}
\newcommand{\eeq}{\end{equation}}
\newcommand{\beqs}{\begin{eqnarray}}
\newcommand{\eeqs}{\end{eqnarray}}

\newcount\colveccount  
\newcommand*\colvec[1]{\global\colveccount#1  \begin{pmatrix} \colvecnext} \def\colvecnext#1{#1 \global\advance\colveccount-1
        \ifnum\colveccount>0 \\ \expandafter\colvecnext
        \else \end{pmatrix} \fi}

\begin{document}

\Large
\centerline{{\bf A phenomenological approach to COVID-19 spread in a population}\\[2ex]}
\centerline{A Thyagaraja}
\centerline{Visiting Research Fellow,  Astrophysics Group, Bristol University, Bristol, UK\\[2ex]}

\begin{center}
\today\\[2ex]
{\bf Summary}
\end{center}

A phenomenological model to describe the Corona Virus Pandemic spread in a given population is developed. It enables
the identification of the key quantities required to form adequate policies for control and mitigation in terms of observable parameters
using the Landau-Stuart equation. It is intended to be complementary to detailed simulations and methods published recently by {\it Ferguson et al,  March 16, (2020)}. The results suggest that the initial growth/spreading rate $\gamma_{c}$ of the disease, and the fraction of infected persons in the population $p_{i}(0)$ can be used to define a ``retardation/inhibition co-efficient''  $k_{*}$, which is a measure of the effectiveness of the control policies adopted. The results are obtained analytically and numerically using a simple Python code. The solutions provide both qualitative and quantitative information. They substantiate and justify two basic control policies enunciated by WHO and adopted in many countries: a) Systematic and early intensive testing individuals for covid-19  and b) Sequestration policies such as ``social distancing'' and ``population density reduction'' by strict quarantining are essential for making $k_{*}>1$, necessary for suppressing the pandemic. The model indicates that relaxing such measures when the infection rate starts to decrease as a result of earlier policies could simply restart the
infection rate in the non-infected population. Presently available available statistical data in WHO and other reports can be readily used to determine the the key parameters of the model. Possible extensions to the basic model to make it more realistic are indicated.

\Large
{\bf 1. Introduction:}\\[2ex]
The Corona Virus [Covid-19] Pandemic is causing devastating health-related and associated economic devastation throughout the World. I was motivated to think about this after reading the data presented by reputable sources \cite{1} and the now-justly famous paper by {\it Ferguson et al, (2019)}\cite{2}. The results obtained by expert epidemiologists depend upon detailed statistical/stochastic modelling based on available data from various countries hardest hit by the Covid-19 Pandemic and on plausible hypotheses. One might term these models as ``microscopic'' in the sense that they take into account of individual interactions and the transmission characteristics of this new disease which appears to have arisen first in China. Although extremely important for national policy-making involving a multitude of factors, it should be possible to understand the principal qualitative properties of the effects of this pandemic. It is therefore important to have some model which is {\bf complementary} to the detailed dynamics of the disease epidemiology which helps in accurately parametrising the overall statistical numbers available.
 
Motivated by the considerations stated in brief above and to aid understanding many confusing and often contradictory statements made by authorities to the public, I have formulated in this Note, a simple, phenomenological mathematical model based on the well-known Landau-Stuart equation\cite{3,4}. This equation has been used in many fields, in ``predator/prey'' models of Lotka-Volterra and notably, in fluid/plasma turbulence. The purpose of this Note is to demonstrate its usefulness in connection with the spread of Covid-19 in any sufficiently large population of uninfected persons.\\[2ex]
{\bf 2. Model assumptions and formulation:}\\[2ex]
A population of {\Red $N_{0}$} individuals is assumed; {\Red $t$} denotes the time.  As a typical example, consider a 
region (for example, an isolated island, country or continent)
in the World with a population of 1 million persons ($N_{0}=10^{6}$). 

Let {\Red $n_{i}(t)$} refer to the number of {\bf infected} individuals in the population. The number of {\bf non-infected}
individuals is denoted by {\Red $n_{w}(t)$}. For the present, we shall take {\Red $N_{0}=n_{i}(t)+n_{w}(t)$}, and
assume no births, deaths or in/out fluxes of people to the population. These additional, more realistic features can also be taken into account within the spirit of the model. 

Consider the spread of an epidemic which does not kill the infected persons, and which does not re-infect a person already infected who recovers from the illness.
 We also assume that initially, there is a ``seed population'' of infected individuals:{\Red $n_{i}(0)\ll N_{0}$; in our example, we could take $n_{i}(0)=10$}. It is assumed that the infection spreads through the population with a {\bf constant growth rate}, {\Red $\gamma_{c}$}, if no action whatever is taken on the system to constrain the spread of infection. It is then obvious that,
{\Red
\begin{eqnarray}
\label{eq.1}
\frac{dn_{i}}{dt}  & = & \gamma_{c}n_{i}
\end{eqnarray}
with the solution, $n_{i}(t)=n_{i}(0)e^{\gamma_{c}t}$ and $n_{w}(t)=N_{0}-n_{i}(t)$}
It follows that the entire population will be infected in the time:
{\Red
\begin{eqnarray}
\label{eq.2}
T_{\rm tot}  & = & (\frac{1}{\gamma_{c}})\ln(\frac{N_{0}}{n_{i}(0)})
\end{eqnarray}}
Clearly, at this time we will see that {\Red $n_{i}(T_{\rm tot})=N_{0}$ and $n_{w}=0$}. It is important to note that the {\bf total infection time} defined by {\Red $T_{\rm tot}$} depends {\it both} on the time-scale defined by {\Red $(\frac{1}{\gamma_{c}})$} and on the {\bf initial infection ratio}, {\Red $\frac{n_{i}(0)}{N_{0}}$}. The first factor is determined both by the infectivity of the virus and the properties of the population considered: for example how dense it is and how it is distributed in space, which is assumed to be ``connected'' in the usual topological sense. In the example, it would be ``simply connected''. The second factor depends upon the relative number of infected persons who were present in the population at the start.

How may one ``mitigate/control'' this naturally exponential (sometimes called ``explosive'') spread of infection? We might attempt to control this rate by testing to statistically estimate the value of {\Red $n_{i}(t)$} once the infection starts and attempt to suppress it by various means, such as ``social distancing'' or ``strictly quarantining''- meaning that we subdivide the population into ``disconnected groups''  which are closed off to all movements of people across their boundaries, so that the spread is limited to disconnected subsets of the whole area.  
This can be simply represented in the present phenomenological model. We might, for example, add a new term in the basic Eq.(\ref{eq.1})
and consider,
{\Red
\begin{eqnarray}
\label{eq.3}
\frac{dn_{i}}{dt}  & = & \gamma_{c}[1-kn_{i}]n_{i}
\end{eqnarray}
}
where {\Red $k$} is a constant (in this simplest model) which represents the {\bf efficacy} with which we estimate the size of {\Red $n_{i}(0)/N_{0}$} and put in measures to reduce the speed of the spreading represented by the spreading rate constant 
{\Red $\gamma_{c}$}. Thus, we are attempting to {\bf retard/suppress} the spreading by reducing(hopefully eliminating) contacts between the infected and the uninfected, by various methods like ``physical/social distancing'' or ``stay-in-place'' policies. This will obviously depend upon 
{\Red $n_{i}/N_{0}$}, since it is this fraction that we normally use to increase the ``retardation'' of the unstable growth of the infection. In the present model, we {\bf assume} that $k$ is a constant. More generally, it could in principle also be a function of  {\Red $n_{i}(t)/N_{0}$} and possibly of  time. Note that when {\Red $n_{i}$} is very small (this means, in practice, {\Red $n_{i}\ll N_{0}$}), the disease will initially grow exponentially with time.

Eq.(\ref{eq.3}) is solved {\em exactly}[see Appendix for detailed mathematical analysis] for arbitrarily specified  values of 
{\Red $\gamma_{c},k, N_{0, }n_{i}(0)$ to get $n_{i}(t)/N_{0}$} of all values of $t$.  Numerical solutions for a typical example which promotes a more detailed, quantitative understanding are also presented and discussed.

Before examining the quantitative implications of the model, it is instructive to consider some of its qualitative properties.\\[2ex]
\newpage
{\bf 3. Qualitative considerations:}\\[2ex]
The Landau-Stuart Eq.(\ref{eq.3}) has some interesting properties which can be inferred from its mathematical form. Some of these also reveal the idealisations in the basic assumptions which lead to it.
\begin{enumerate}
\item
Evidently, if $n_{i}(0)=0$, we see that it will remain zero for all $t$. This property is ``trivial'' in the sense, the model does not describe the ``spontaneous'' appearance of the pandemic. This property is also shared by its much simpler linear ``progenitor'', Eq.(\ref{eq.1}).
\item
If we take no action, obviously {\Red $k=0$} and we have ``explosive/exponential spread'', since then Eq.(\ref{eq.3}) 
reduces to Eq.(\ref{eq.1}).
\item
If {\Red $k>0$, and $n_{i}(t)<1/k$}, the equation says that {\Red $n_{i}(t)$} will continue to grow in time, although not at the rate, 
{\Red $\gamma_{c}$} of ``uncontrolled spreading''. It is readily seen [from Eq.(\ref{eq.3})]  that, as {\Red $kn_{i}(t)$} grows from some initially small {\Red $kn_{i}(0)$},
the ``effective growth rate'' [ie,{\Red  $\gamma_{c}(1-kn_{i})$}] keeps reducing and would become zero at a time {\Red $T_{\rm max}$ when $kn_{i}(T_{\rm max})=1$}. After this, {\Red $n_{i}(t)$} is at its maximum value, {\Red $n_{\rm max}$}. We shall show in the Appendix that {\Red $T_{\rm max}$} is actually infinity! What this means is that as {\Red $t$} increases, {\Red $n_{i}(t)$} always remains smaller than {\Red $n_{\rm max}=1/k$}, but approaches it  arbitrarily closely from below. 
\item
The observations above imply that we should make the constant (in this simple model) {\Red $k$} as large as we can by implementing adequate ``infection-retardation''  control measures. Evidently, what matters here is the fact that accuracy of testing is likely to be decided by the {\bf probable number of people} who developed the illness at {\Red $t$}. This is represented in the model by the ratio, {\Red $n_{i}/N_{0}$}. 

Thus, we set,
{\Red $k=k_{*}/N_{0}$, where $k_{*}$} is a pure number [it seems reasonable to call this the {\bf infection retardation coefficient}]. Then, {\Red $kn_{i}=k_{*}(\frac{n_{i}}{N_{0}}), k_{*}=kN_{0}$}. 
We may then write the {\bf effective growth/spreading rate}={\Red $\Gamma_{\rm eff}(t)$} in the form:
{\Red
\begin{eqnarray}
\label{eq.4}
\Gamma_{\rm eff}(t)
      & = & \gamma_{c}[1-k_{*}(\frac{n_{i}}{N_{0}})]
\end{eqnarray}}
Note that unlike {\Red $\gamma_{c}$}, the effective growth rate depends upon the relative size of the infected cohort with reference to the initial total population, and is clearly a function of time. Since, we are usually interested in mitigating (rather than aggravating!) the growth rate, our efforts must try to make {\Red $k_{*}>0$.  

If $k_{*}< 1$}, it is easy to see that although the growth will slow as time goes on, 
{\Red $\Gamma_{\rm eff}(t)>(1-k_{*})>0$ even when $n_{i}(t)=N_{0}$}, and everyone in the population will have been infected. Hence, we require, for significant retardation, {\Red $k_{*}>1$}. 

Provided we can formulate controls to ensure {\Red $k_{*}>1$}, we can derive an interesting consequence: for very long times, the total number infected is predicted by Eq.(\ref{eq.4}) to be, slightly less than, {\Red $n_{i}^{\rm max}=\frac{N_{0}}{k_{*}}<N_{0}$}. Thus, this model predicts that, 
{\Red$$n_{i}^{\rm max}=\frac{N_{0}}{k_{*}}$$}
 is an upper limit (as $t\rightarrow \infty$) to the number of persons who can get the infection provided {\Red $k_{*}$} is kept constant in time. 
 
 It is easy to see that if at any time after the start of applying {\Red $k_{*}$} we ``relax'' our controls and allow it to decrease below unity, the disease will start to spread again into the remaining non-infected population. This raises important questions for policy makers. The model does not take account of {\bf asymptomatic infectors}. It is at present not clear from the available data to know if people cured of the illness can infect the non-infected and whether asymptomatic disease transmission really exists.
\item
The ``spreading equation'' Eq.(\ref{eq.3}) can be written in a more understandable form. It is convenient to introduce 
{\Red $p_{i}(t)=\frac{n_{i}}{N_{0}}$} and divide the equation through by {\Red $N_{0}$}, to obtain,
{\Red
\begin{eqnarray}
\label{eq.5}
\frac{dp_{i}}{dt}  & = & \gamma_{c}[1-k_{*}p_{i}]p_{i}
\end{eqnarray}
}
This form immediately leads to [provided {\Red $k_{*}p_{i}(0)<1$}]:
{\Red $p_{i}(t) < \frac{1}{k_{*}}$, $\frac{1}{k_{*}}=p_{i}^{\rm max}$, for all $t\geq 0$}. Note that 
{\Red $p_{i}^{\rm max}$} depends {\bf only} upon our ``retardation constant'', {\Red $k_{*}$}. In particular, it does not depend upon the 
growth rate constant, {\Red $\gamma_{c}$, nor, remarkably on $N_{0}$ or $p_{i}(0)$!}
\item
In the Appendix we show using Eq.(\ref{eq.5}) that the maximum value of the {\bf instantaneous rate of growth} [defined by,
{\Red $\frac{dp_{i}}{dt}$}] occurs when {\Red $k_{*}p_{i}=1/2$}, namely, when {\Red $p_{i}(t) =\frac{1}{2k_{*}}$}. The exact solution of the equation gives the time when this happens.
\end{enumerate}
{\bf 4. Discussion of exact and numerical solutions:}\\[2ex]
The exact solution for {\Red $p_{i}(t)$ [equivalent to $n_{i}(t)$ in this model]} shows that the {\bf instantaneous infection spread rate}, is an interesting function:
{\Red
$$\dot{p}_{i}=\frac{p_{i}^{\rm max}c_{0}\gamma_{c}\exp{\gamma_{c}t}}{(1+c_{0}\exp{\gamma_{c}t})^{2}}$$
}
This can be written is a more illuminating form:
{\Red 
\begin{eqnarray}
\label{eq.6}
\dot{p}_{i}(t)
          & = & (\gamma_{c}p_{i}^{\rm max})
           \frac{1}{c_{0}}\left [1-\frac{1}{(1+c_{0}e^{\gamma_{c}t})}]^{2}\right ]
\end{eqnarray} 
}

The first factor on the right is constant in time, governed only by {\Red $\gamma_{c}/k_{*}$}. The second and third terms depend upon,  {\Red $k_{*}, c_{0}, \gamma_{c}t$} as well (since, to a good approximation, 
 {\Red $c_{0} \approx x_{0}=x(0)=k_{*}p_{i}(0)$; this follows from, $p_{i}(0) \ll 1$}). Notably, it does {\bf not} depend upon {\Red $N_{0}$}. The time-dependence is only through the product, {\Red $\gamma_{c}t$}.

We can discuss qualitatively the course of this key function in time: At {\Red $t=0$}, it is easily shown that the time-dependent term within parenthesis equals {\Red $c_{0}^{2}$} to an excellent approximation. Then, we see that  {\Red $\dot{p}_{i}(0)= (\gamma_{c}p_{i}^{\rm max})c_{0}=\gamma_{c}p_{i}(0)$}, consistent with Eq.(\ref{eq.5}) when {\Red $k_{*}p_{i}(0)$} is very small compared to unity.  As {\Red $t$} increases, this increases and reaches a maximum when {\Red $p_{i}=p_{i}^{\rm max}/2$}. We can verify that at this time [{\Red $t_{\rm peak}=\ln(\frac{p_{i}^{\rm max}}{p_{i}(0)})$}], Eq.(\ref{eq.5}) implies: {\Red $\dot{p}_{i}^{\rm max}=\gamma_{c}/4$}.

While the exact solution is useful for demonstrating theoretical features of the model [such as relating the basic parameters to epidemiologically important objects like {\Red $p_{i}^{\rm max}$} and the fact that the ``invariant'' profile of the rate of increase], it is useful to solve Eq.(\ref{eq.5}) numerically for a typical case. 

In the following, the Example introduced earlier
will be elaborated: thus, it has already been stated that $n_{i}(0)/N_{0}=p_{i}(0)=1.0^{-5}$ (ie, 10 infected individuals in a million). If we take the unit of time as a day, anecdotal evidence suggests that in two days, the disease will spread to 2 or 3 individuals. For this reason,  {\Red $\gamma_{c}=0.5$} is assumed. The calculation is carried out for a period of 50 days. Keeping these parameters {\it fixed} the following values of $k_{*}=0,0.5,1.0,1.5,5.0$ are considered and two interesting quantities, {\Red $p_{i}$ and $\frac{dp_{i}}{dt}$} as functions of $t$ are plotted in Figs. 1-5.
\begin{figure}[h!]
   \centering
   \begin{subfigure}[b] {0.4\linewidth}
      \includegraphics[width=\linewidth] {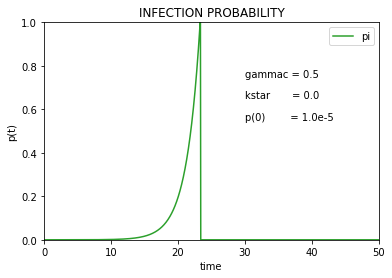}
      \caption{$p_{i}$ vs. $t$}
    \end{subfigure}
       \begin{subfigure}[b] {0.4\linewidth}
      \includegraphics[width=\linewidth] {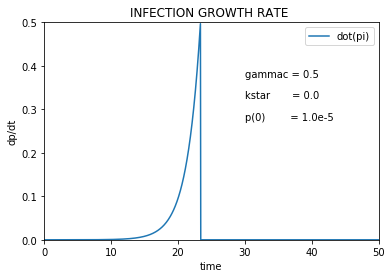}
      \caption{$\frac{dp_{i}}{dt}$ vs. $t$}
    \end{subfigure}
\caption{ \Large Infection Probability and Infection Rate as functions of time(in days): $k_{*}=0$}
\end{figure}

Figure 1 shows what happens when $k_{*}=0.0$. Fig. 1A shows the rapid rise of {\Red $p_{i}(t)$}. It is evident that 
for the parameters assumed in the calculation, in about 23 days, {\Red$n_{i}(t)=N_{0}$}. This is {\Red $T_{\rm tot}$}. After this,  ``infection probability'' does not make sense and the calculation is stopped. Fig. 1B shows a similar behaviour of the infection growth rate. In this case, the two rates are simply proportional to each other as shown by Eq.(\ref{eq.5}). The proportionality is simply $\gamma_{c}=1/2$. This is a {\it reference calculation} which shows the effect of simply ignoring the disease.
\begin{figure}[h!]
   \centering
   \begin{subfigure}[b] {0.4\linewidth}
      \includegraphics[width=\linewidth] {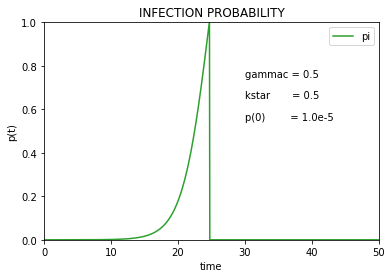}
      \caption{$p_{i}$ vs. $t$}
    \end{subfigure}
       \begin{subfigure}[b] {0.4\linewidth}
      \includegraphics[width=\linewidth] {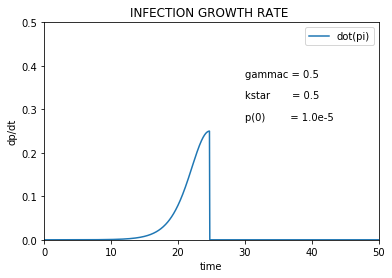}
      \caption{$\frac{dp_{i}}{dt}$ vs. $t$}
    \end{subfigure}
\caption{ \Large Infection Probability and Infection Rate as functions of time(in days): $k_{*}=0.5$}
\end{figure}

Figure 2 demonstrates the effect of choosing {\Red $k_{*}=1/2$}. As discussed earlier, this is less than the `critical value' {\Red $k_{*}=1$}.
It is seen that {\Red $p_{i}=1$} around 24 days. Remarkably, this sub-critical value of the co-efficient only lengthens the time-scale of the total infection of the population of 1 million individuals (starting with 10 at the start) slightly. 
However,
the infection rate is lower by 50\%, relative to the ``do nothing'' scenario. This illustrates that half-measures will not be effective in really controlling the disease and protecting a significant fraction of the population. As before, the calculation stops
at the time not substantially different from {\Red $T_{\rm tot}$}.  

Figure 3 shows that when {\Red $k_{*}=1$, $p_{i}$} keeps rising monotonically and {\Red $n_{i}(t)$} always remains less than {\Red $N_{0}$}. 
By about 
40 days the population does get fully infected. In actual fact, {\Red $p_{i}(t)<1$}, and approaches this value as a limit. This is demonstrable from Eq.(\ref{eq.5}). In this case, we see that {\Red $\frac{dp_{i}}{dt}$} has a ``bell-like'' shape with a maximum at 1/8. Contrary to popular belief, the curve is {\bf not} a Gaussian, nor is it really symmetrical about the maximum. It is, in fact, correctly represented by the formula, Eq.(\ref{eq.6}). The maximum rate of infection occurs around 24 days. 

Figure 4 shows the the effect of choosing {\Red $k_{*}=3/2$}, a ``super-critical'' value. This time, {\Red $p_{i}$} adopts the classic S [called a ``saturation curve'']
shape as in the critical case. However, {\Red $p_{i}^{{\rm max}}=1/k_{*}$}, as predicted theoretically from exact solution of 
Eq.(\ref{eq.6}). Thus this policy certainly shields 33\% of the initial population from  the illness. Furthermore, the infection rate curve in Fig. 4B, is certainly ``flattened'' as compared with Fig. 3B. However, the area under it is {\bf not equal} to the curve in Fig. 2B. In actual fact, it must be 33\% smaller as also predicted by the analysis.
\begin{figure}[h!]
   \centering
   \begin{subfigure}[b] {0.45\linewidth}
      \includegraphics[width=\linewidth] {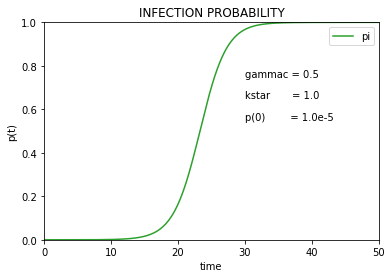}
      \caption{$p_{i}$ vs. $t$}
    \end{subfigure}
       \begin{subfigure}[b] {0.45\linewidth}
      \includegraphics[width=\linewidth] {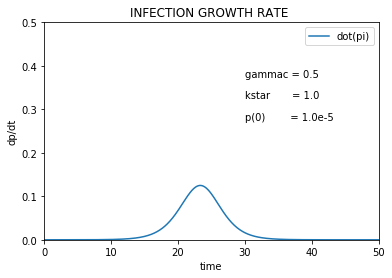}
      \caption{$\frac{dp_{i}}{dt}$ vs. $t$}
    \end{subfigure}
\caption{ \Large Infection Probability and Infection Rate as functions of time(in days): $k_{*}=1.0$}
\end{figure}
\begin{figure}[h!]
   \centering
   \begin{subfigure}[b] {0.45\linewidth}
      \includegraphics[width=\linewidth] {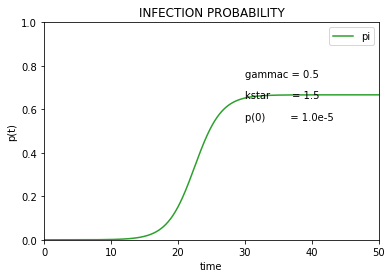}
      \caption{$p_{i}$ vs. $t$}
    \end{subfigure}
       \begin{subfigure}[b] {0.45\linewidth}
      \includegraphics[width=\linewidth] {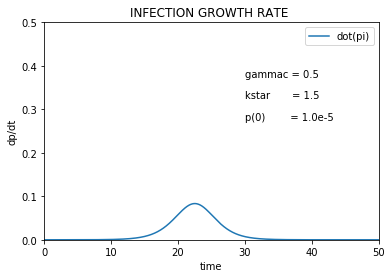}
      \caption{$\frac{dp_{i}}{dt}$ vs. $t$}
    \end{subfigure}
\caption{ \Large Infection Probability and Infection Rate as functions of time(in days): $k_{*}=1.5$}
\end{figure}
\begin{figure}[h!]
   \centering
   \begin{subfigure}[b] {0.45\linewidth}
      \includegraphics[width=\linewidth] {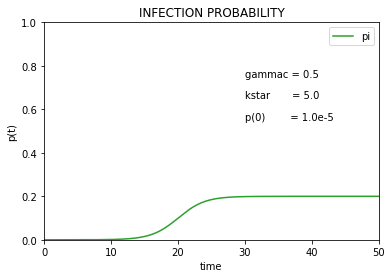}
      \caption{$p_{i}$ vs. $t$}
    \end{subfigure}
       \begin{subfigure}[b] {0.45\linewidth}
      \includegraphics[width=\linewidth] {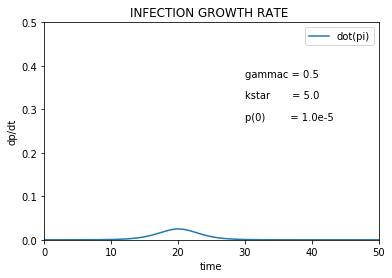}
      \caption{$\frac{dp_{i}}{dt}$ vs. $t$}
    \end{subfigure}
\caption{ \Large Infection Probability and Infection Rate Infection Rate as functions of time(in days): $k_{*}=5.0$} 
\end{figure}

Figure 5 shows that by having a relatively large {\Red $k_{*}=5.0$}, one can shield nearly 80\% of population and really flatten the infection rate curve. This should not cause any surprise since {\Red $k_{*}p_{i}$} reaches unity by about 25 days. After that, the effective growth rate, {\Red $\Gamma_{\rm eff}<0$}.

This concludes the discussion of the analytical and numerical results of the model presented. In the next Section some 
possible extensions and applications of similar, but more realistic models in understanding the variety of data from around the world and producing a coherent explanation of the significant variations among different countries are suggested.\\[2ex]
{\bf 5. Conclusions:}\\[2ex]
The model presented can be extended and generalised in several ways to take better account of the observations of Covid-19. 
Some of these will now be briefly discussed. The model, as stated, does not account for {\bf mortality} as a result of infection. 
To represent this, one needs to introduce, the probability that an individual would die after infection. If we set {\Red$n_{d}(t)$} for
number of deaths {\bf due to the infection}, and assume a mean probability of deaths equal to 
{\Red $n_{d}/n_{i}(t)=r_{d}\leq 1$}, we can calculate {\Red $n_{d}$} after obtaining {\Red $n_{i}$} as described earlier. It is known that this probability varies with many factors and is certainly not the same for all age groups\cite{1}. We can follow the distribution of ({\Red $p_{i}(t)$}) among different age groups by introducing the distribution function, {\Red $f_{i}(y,t)$} where, {\Red $y$} is the age of the person. It is clear that ,
{\Red $$n_{i}(t)=\int_{0}^{Y}f_{i}(y,t)dy$$}
 where {\Red $Y$} is the largest age of a person in the population. We must develop equations for 
{\Red $f_{i}(y,t)$} similar in form to Eq.(\ref{eq.3}). If all other deaths, births etc are ignored and we consider times shorter than a year,
the total population {\Red $N_{0}$}, will no longer be constant in time, but decrease. If the mortality {\Red $r_{d}$} of a person depends upon age but not on the time, we estimate the number of deaths in time {\Red $t$} by, 
{\Red $$n_{d}(t)=\int_{0}^{Y}r_{d}(y)f_{i}(y,dt)$$}
In principle, it is not difficult to set up equations corresponding to Eq.(\ref{eq.3}) already data exists relating to {\Red $r(d)$}, and is bound to be improved as experts in the field have longer to validate statistical results 

It is also relatively straight forward to account for possible ``seasonal'' variations of  {\Red $\gamma_{c}$}, in the case of the simple model described by Eq.(\ref{eq.3}).  Thus, when {\Red $\gamma_{c}(t)$} is taken to be of the form, 
{\Red $$\gamma_{c}(t)=\gamma_{c0}(1+g_{1}\sin \omega t+..)$$}
which is a typical periodic function (which equals {\Red $\gamma_{c0}$} of time with period {\Red $2\pi/\omega$}, Eq.(\ref{eq.3}) takes the form,
{\Red $$\frac{dn_{i}}{dt} =  \gamma_{c}(t)[1-kn_{i}]n_{i}$$}
To solve this, one introduces a new `time-like' variable, {\Red $\tau=\int_{0}^{t}\gamma_{c}(u)du$}. In terms of this new independent variable, the modified ``spreading equation'' takes the form:
{\Red $$\frac{dn_{i}}{d\tau} =  \gamma_{c}(\tau)[1-kn_{i}]n_{i}$$}
This standard change of variable can be actually implemented in principle (numerically, if necessary) once the seasonal variation coefficients,
{\Red $\gamma_{c}^{0}, g_{1},..$} are known. 

What of cases where the `retardation coefficient', {\Red $k_{*}$} {\it itself} varies with time? This could occur during the pandemic as policy changes are implemented over time. It is easy to see that if {\Red $k_{*}(t)$} is known, we can simply define {\Red $k_{*}(t)p_{i}=p_{*}(t)$} and transform  Eq.(\ref{eq.9}) to the form,
{\Red $$ \frac{dp_{*}}{dt}-(\frac{p_{*}}{k_{*}})\frac{dk_{*}}{dt}=\gamma_{*}k_{*}[1-p_{*}]$$}
It is interesting to note that if the control measures are programmed to increase, as the infection starts to spread (making {\Red $k_{*}(t)$} an increasing function of {\Red $t$}), the rate of increase can be `flattened' even faster. Conversely, if 
{\Red $k_{*}$} decreases as a function of time, the rate of spreading will necessarily increase.
If the variation with time is complicated (such as a `step function'), the equation can still be solved over every 'interval of constancy' and the overall solution obtained numerically by continuing the solutions across the boundary points of each interval. 

As is characteristic of any phenomenological model, the parameters such as {\Red $\gamma_{c}, k_{*}, p_{i}(0)$} have to be empirically obtained. Thus available data can be used to estimate such model parameters. More detailed epidemiological statistical models will certainly be needed to obtain such parameters from `first principles'. An analogy from Physics may be relevant here: just as Thermodynamics describes matter in a gross, phenomenological
manner and has to be supplemented by detailed molecular models to relate its results to fundamental physical principles (eg. Classical or Quantum Statistical Mechanics/Kinetics). The model described in this Note is best at giving a relatively simple ``macro-level'' description of a very complex epidemiological dynamical system represented by the Covid-19 pandemic.\\[2ex] 
{\bf Appendix:}\\[2ex]
The analysis of Eq.(\ref{eq.5}) is presented here.

It is convenient to introduce the new variable, {\Red $x(t)=k_{*}p_{i}(t)=\frac{p_{i(t)}}{p_{i}^{\rm max}}$}, and a
new ``relative time'', {\Red $\tau(t)=\gamma_{c}t$}. Thus {\Red $\tau$} is the measure of time in terms of the ``growth time'' {\Red $1/\gamma_{c}$}, characteristic of the disease. It is easily related to the observed rates of the disease (with some errors due to inadequate testing etc). We find then that Eq.(\ref{eq.5}) becomes, when written in the new variables,
{\Red
\begin{eqnarray}
\label{eq.7}
\frac{dx}{d\tau}  
            & = & (1-x)x \\
\tau+C
            & = & \int \frac{dx}{x(1-x)} \nonumber 
\end{eqnarray}}
leading to,
{\Red
\begin{eqnarray}
\frac{x}{1-x}
            & = & c_{0}\exp{\tau} \nonumber
 \end{eqnarray}}
The initial condition, {\Red $x_{0}=x(0)$ gives $c_{0}=\frac{x_{0}}{1-x_{0}}$}. 
We obtain from this,
{\Red
\begin{eqnarray}
\label{eq.8}
x(\tau)
            & = & 1-\frac{1}{1+c_{0}\exp{\tau}}
\end{eqnarray}}
We can express this in terms of the original variables to get a feel for the character of the solution and some of its key properties.
{\Red 
\begin{eqnarray}
\label{eq.9}
p_{i}(t)
         & = & p_{i}^{\rm max}\left [1-\frac{1}{1+c_{0}\exp{\gamma_{c}t}}\right]
\end{eqnarray}}

Thus, it is clear that at 
{\Red $t=0$, $p_{i}(0)=p_{i}^{\rm max}x_{0}$ and as $t \rightarrow \infty$, $p_{i}(t) \rightarrow p_{i}^{\rm max}=\frac{1}{k_{*}}$}. 
Since {\Red $p_{i}(t)$} is manifestly increasing function of 
{\Red $t$}, we see that  {\Red $\frac{1}{k_{*}}p_{i}(t)< p_{i}^{\rm max}<1$}.

We next examine when the rate of new infections defined as {\Red$\frac{dp_{i}}{dt}$} is a maximum. Evidently, this will occur
at a time {\Red $\tau_{\rm peak}$}, when 
{\Red $$\frac{d^{2}p_{i}}{dt^{2}}=0$$}
From Eq.(\ref{eq.9}) this happens when $t$ satisfies,
{\Red
\begin{eqnarray}
\label{eq.10}
\frac{d}{dt}[\frac{c_{0}\gamma_{c}\exp{\gamma_{c}t}}{(1+c_{0}\exp{\gamma_{c}t})^{2}}]
           & = & 0 
\end{eqnarray}
which implies the relation, $1-2\frac{c_{0}\exp{\gamma_{c}t}}{1+c_{0}\exp{\gamma_{c}t}}=0$,  {\bf precisely the condition}}
{\Red $$p_{i}(t)=p_{i}^{\rm max}/2$$}
predicted from Eq.(\ref{eq.3}).

Eq.(\ref{eq.9}) then leads to:
{\Red
\begin{eqnarray}
\label{eq.11}
\tau_{\rm peak}
     & = & \frac{1}{\gamma_{c}}\ln(\frac{1}{c_{0}})
\end{eqnarray}
Since $c_{0}~x_{0}$ when $x_{0}\ll 1$, 
$$\tau_{\rm peak}= \frac{\ln(1/x_{0})}{\gamma_{c}}$$
to an excellent approximation.}
\bibliographystyle{plain} 

\end{document}